\documentclass[aps,prd,reprint,superscriptaddress,nofootinbib,longbibliography]{revtex4-2}
\usepackage{graphicx}
\usepackage{amsmath}
\usepackage{booktabs}
\usepackage{multirow}
\usepackage[colorlinks=true,citecolor=blue,urlcolor=cyan,linkcolor=black]{hyperref}
\usepackage{fontawesome5} 
\usepackage{amsfonts}
\usepackage{simplewick}
\usepackage{bm}

\usepackage{tabularx}
\newcolumntype{C}{>{\centering\arraybackslash}X}
\newcolumntype{R}{>{\raggedleft\arraybackslash}X}

\def\spinup{\partial\kern-0.3em\raise0.42ex\hbox{\tiny\textbackslash}}
\def\spindown{\overline{\partial\kern-0.3em\raise0.42ex\hbox{\tiny\textbackslash}}}

\def\hv{{\hat v}}

\def\Cov{\mbox{Cov}}

\def\N{{\mathcal N}}

\newcommand{\vpot}{\Upsilon}

\newcommand{\hpi}{\widehat{\pi}}
\newcommand{\Pge}{P_{ge}}
\newcommand{\bksz}{A_{ge}}

\newcommand{\NactFreqMainw}{six}

\newcommand{\NcapdiffMainw}{four}
\newcommand{\NnullAll}{46}

\newcommand{\NnullExtDnw}{ten}
\newcommand{\NnullExtNight}{13}

\newcommand{\NnullMainDn}{10}

\newcommand{\NnullMainNight}{13}

\newcommand{\NnullNight}{26}

\newcommand{\Npairsw}{nine}

\newcommand{\NpassExtDnw}{nine}

\newcommand{\NsptFreqMainw}{three}

\newcommand{\Ntwotwentymainw}{ten}

\newcommand{\hbksz}{\hat{A}_{ge}}
\newcommand{\hvr}{{\hat v}_r}

\usepackage[normalem]{ulem}
\usepackage[dvipsnames]{xcolor}

\def\bx{{\boldsymbol{x}}}
\def\bell{{\boldsymbol{\ell}}}

\newcommand{\btheta}{{\boldsymbol{\theta}}}

\newcommand{\bk}{\boldsymbol{k}}
\newcommand{\bl}{\boldsymbol{l}}
\newcommand{\be}{\begin{eqnarray}}

\newcommand{\ee}{\end{eqnarray}}

\definecolor{colorA}{HTML}{1E90FF}
\definecolor{colorB}{HTML}{228B22}
\definecolor{colorC}{HTML}{FF7F00}
\definecolor{colorD}{HTML}{4B0082}
\definecolor{colorE}{HTML}{B22222}

\definecolor{lgreen}{HTML}{32CD32}
\definecolor{lgray}{HTML}{D3D3D3}
\definecolor{dblue}{HTML}{1E90FF}
\definecolor{orange}{HTML}{FF4500}
\definecolor{indigo}{HTML}{4B0082}

\definecolor{teal}{HTML}{008080}
\definecolor{firebrick}{HTML}{B22222}
\definecolor{salmon}{HTML}{FA8072}
\definecolor{darkgreen}{HTML}{006400}

\newcommand{\perimeter}{Perimeter Institute for Theoretical Physics, 31 Caroline St N, Waterloo, ON N2L 2Y5, Canada}

\graphicspath{ {plots/} }
\begin{document}
\raggedbottom 

\title{The galaxy--electron cross-spectrum with ACT, SPT-3G and DESILS LRGs}

\author{Selim~C.~Hotinli}
\email{shotinli@perimeterinstitute.ca}
\affiliation{\perimeter}

\date{\today}

\begin{abstract}
The galaxy--electron cross-power spectrum $\Pge(k)$ measures the small-scale distribution of gas around galaxies, the quantity reshaped by feedback.
We measure $\Pge$ from the kinetic Sunyaev--Zel'dovich (kSZ) signal of the photometric DESI Legacy Imaging Surveys luminous red galaxies, cross-correlating a velocity-weighted galaxy momentum template with the CMB temperature from ACT~DR6 and, for the first time, SPT-3G.
The absolute normalization of the cross-spectrum is fixed by a surrogate-field Monte Carlo that propagates the survey mask, lightcone evolution, and photometric-redshift weights, converting the measured pseudo-spectrum into $\Pge(k)$ in physical units.
We detect $\Pge$ at $11.7\sigma$ with ACT and $8.4\sigma$ with SPT-3G, and at $12.3\sigma$ combining SPT-3G with the ACT northern cap, which it does not overlap.
The two surveys have independent optics, beams, filtering and noise and share about two thirds of the SPT-3G galaxies; their amplitudes agree channel by channel, the largest difference being $0.7\sigma$.
All $\NnullNight$ null tests on the adopted night maps are consistent with zero signal, among them the \NactFreqMainw{} ACT frequency differences, the \NsptFreqMainw{} SPT-3G frequency differences, and the \NcapdiffMainw{} ACT cap differences of the main selection.
Turning the amplitude into a gas fraction requires the host halo masses of the sample, which the cross-spectrum does not itself measure; taking them from its measured number density and from a published CMB-lensing calibration of the same galaxies, the gas within the virial radius is $25$--$33\%$ of the cosmic baryon budget of the halo.
\end{abstract}

\maketitle

\section{Introduction}\label{sec:intro}

A large fraction of the baryons in the Universe resides in diffuse, ionized gas in and around dark-matter halos, redistributed by the feedback processes that accompany galaxy formation, among them active galactic nuclei and supernovae~\cite{Dave:2000vh,2004ApJ...616..643F,Cen:2006by,Bregman:2007ac,2012ApJ...759...23S,Nicastro:2018eam,Macquart:2020lln}.
This gas is faint and hard to observe directly, and its distribution on halo scales is among the most uncertain ingredients in the theory of galaxy formation, as well as one of the leading systematics for upcoming weak-lensing cosmology, which must marginalize over how feedback rearranges matter on small scales~\cite{vanDaalen:2011xb,Chisari:2019tus,Amon:2022azi,DES:2024iny,McCarthy:2024tvp,Siegel:2025ivd,Sunseri:2025ilf}.

One observable that carries the distribution of gas around galaxies is the galaxy--electron correlation, described in Fourier space as the cross-power spectrum $\Pge(k)$. On small scales, this correlation is set by how the gas is arranged around the host halos and galaxies, and its shape there measures how far the gas extends beyond the halo and what fraction of the baryons remains within it, the two quantities on which hydrodynamical simulations differ from one another~\cite{vanDaalen:2011xb,Chisari:2019tus,Dave:2019yyq,Nelson:2018uso,Schaye:2023jqv,Hadzhiyska:2026aeo}.

The kinetic Sunyaev–Zel'dovich (kSZ) effect~\cite{Sunyaev1980} provides a route to $\Pge$ from the CMB.
CMB photons scattering off free electrons are Doppler-shifted by the motion of the gas, so the kSZ temperature is the electron density times the radial velocity, integrated along the line of sight.
A galaxy survey isolates this signal by weighting galaxies by their reconstructed large-scale radial velocities to form a momentum template and cross-correlating it with the CMB temperature; in the squeezed limit, that cross-spectrum is proportional to $\Pge(k)$~\cite{Smith:2018bpn}.

Schaan, Ferraro \emph{et al.} first measured the gas density profile to be more extended than the dark matter, stacking ACT maps on the spectroscopic BOSS CMASS galaxies~\cite{ACTPol:2015teu,AtacamaCosmologyTelescope:2020wtv,Amodeo:2020mmu}, and stacking on the spectroscopic DESI~Y1 luminous red galaxies (LRGs) gives consistent profiles~\cite{RiedGuachalla:2025byu}.
Pairwise-momentum statistics gave the first kSZ detections~\cite{Hand2012,Chen2022,Hadzhiyska:2025egz}.
Fourier-space estimators have measured $\Pge(k)$ itself, first explored by~\cite{Sugiyama:2017uvr} and formalized by~\cite{Smith:2018bpn}, applied recently with ACT~\cite{Qu:2026zyh,Hadzhiyska:2026uyq} and in the closely related pseudo-$C_\ell$ construction of~\cite{Harscouet:2025pwl,Wayland:2025fte}; they favor feedback efficiencies exceeding those in the Battaglia profile~\cite{Qu:2026zyh}, for example.

The conclusive kSZ measurements of the gas around photometric LRGs are the work of Hadzhiyska, Ferraro \emph{et al.}, who stacked ACT~DR6 maps on the DESI Legacy Imaging Surveys LRG (DESILS-LRG) samples we analyze in this work, and also found strong evidence that the gas is significantly more spread out than the dark matter~\cite{Hadzhiyska:2024qsl}.
The same authors calibrated the halo masses of those galaxies from the ACT~DR6 CMB lensing convergence and recovered the full cosmic baryon budget of the halo by around three virial radii~\cite{Hadzhiyska:2025mvt}.

SPT-3G overlaps the DESILS-LRG footprint in the south with deeper maps at every frequency, $5.4$, $4.4$ and $16.2\,\mu\mathrm{K}$-arcmin at $95$, $150$ and $220$\,GHz~\cite{SPT-3G:2026krg}, against $15$, $16$ and $74$ at $90$, $150$ and $220$\,GHz, median over the exposed area, for the ACT~DR6 night coadds~\cite{Naess2025}.
$\Pge$ has not been measured with SPT-3G before, so our measurement provides new evidence for the gas distribution found with ACT on the photometric DESILS-LRG samples~\cite{Hadzhiyska:2024qsl,Hadzhiyska:2025mvt} and on DESI spectroscopic LRGs~\cite{RiedGuachalla:2025byu,Qu:2026zyh,Hadzhiyska:2026uyq}; from a telescope that shares no optics, beams, filtering, or noise with ACT.
Under the footprints and cuts of this analysis, $64\%$ of the SPT-3G galaxies also enters our ACT measurement, so an instrumental systematic in either would show up as a disagreement between them; the remaining galaxies lie south of the ACT coverage, and the ACT northern cap lies outside the SPT-3G field, so each survey also covers sky that the other does not.

We measure $\Pge$ for the DESILS-LRG samples with ACT~DR6 and SPT-3G, quoting the brighter ``main'' selection throughout; the fainter ``extended'' selection is measured the same way and is described in Appendix~\ref{app:extended}.
We report the bandpowers themselves and summarize each measurement by a single amplitude $\bksz$, which represents the level of the measured $\Pge$ relative to a fiducial halo-model template.
The ACT $90+150$\,GHz combination gives $\bksz=0.408\pm0.035$ ($11.7\sigma$, both Galactic caps combined), and the first SPT-3G measurement $\bksz=0.392\pm0.047$ ($8.4\sigma$) at $95+150$\,GHz; combining SPT-3G with the northern ACT cap gives $\bksz=0.416\pm0.034$ ($12.3\sigma$).
The two surveys agree channel by channel, most closely on the southern sky that they share.
Both lie well below the unscaled template, with the ACT combination deviating by about a factor of $2.5$, and every null test on the night maps is consistent with zero signal.
Read as a gas fraction on a halo mass, the amplitude puts $25$--$33\%$ of the cosmic baryon budget of the halo within the virial radius, across both surveys, against $63\%$ for the same profile at $\bksz=1$.

The cross-spectrum determines $\Pge$ only up to how strongly the momentum template responds to the true velocity field, and that response sets the normalization for the estimator we introduce below, and hence sets the scale of every number we quote, $\bksz$ and the gas fraction alike.
Other analyses take it from simulations matched to their sample~\cite{AtacamaCosmologyTelescope:2020wtv, Hadzhiyska:2024qsl, RiedGuachalla:2025byu, Hadzhiyska:2025mvt, Qu:2026zyh, Hadzhiyska:2026uyq}.
Here, we compute it from the survey itself, with a surrogate Monte Carlo that carries Gaussian realizations through the same pipeline that the data go through.
Because we generate the velocity field of each realization, the response is known exactly within it. Therefore, averaging over realizations fixes the normalization without a mock catalog and without leaving anything in the amplitude fit free to absorb it; our values sit in the band simulations provide for comparable samples. 

The catalog handling, the curved-sky pipeline, and the surrogate-field construction build on earlier kSZ velocity-reconstruction analyses~\cite{Hotinli:2025tul,Chaussidon:2026vmn}, in code built on the publicly available {\tt kszx} package~{\href{https://github.com/kmsmith137/kszx}{\faGithub}} and extended in this work.
The small-scale estimator, the surrogate evaluation of its absolute normalization, and the statistical analysis are developed in this paper. We develop the estimator, its surrogate normalization and the statistical analysis in Section~\ref{sec:pipeline}, describe the CMB maps and galaxy catalogs in Section~\ref{sec:data}, present the detections and the null tests in Section~\ref{sec:results}, and conclude in Section~\ref{sec:discussion}.
Three appendices follow.

\section{Preliminaries}

Throughout, we denote three-dimensional comoving positions by $\bx$ and two-dimensional sky directions by the unit vector $\btheta$, with $\hat\bx$ the line-of-sight direction (oriented away from the observer).
The comoving distance to redshift $z$ is $\chi(z)$, and we change variables $\chi\leftrightarrow z$ without comment. We use an effective redshift at the median of our galaxy sample $z_*=0.734$, and comoving distance $\chi_*\equiv\chi(z_*)$, at which we evaluate the small-scale signal. Fourier transforms carry the $(2\pi)^{-3}$ with the $d^3\bk$ integral and no factor with $d^3\bx$.

The kSZ anisotropy is a line-of-sight integral,
\begin{equation}\label{eq:tksz_los}
T_{\rm kSZ}(\btheta) = \int d\chi\,K(\chi)\,v_r(\chi\btheta)\,\delta_e(\chi\btheta)\,,
\end{equation}
where $v_r$ is the radial peculiar velocity, $\delta_e$ the free-electron overdensity, and the kSZ radial kernel is:
\begin{equation}\label{eq:K_def}
K(z) \equiv -T_{\rm CMB}\,\sigma_T\,n_{e0}\,x_e(z)\,e^{-\tau(z)}\,(1+z)^2\,,
\end{equation}
with $T_{\rm CMB}$ the mean CMB temperature, $\sigma_T$ the Thomson cross-section, $n_{e0}$ the mean electron density today, $x_e$ the ionization fraction, and $\tau$ the optical depth.
We set $x_e\,e^{-\tau}\simeq 1$, and $K_*\equiv K(z_*)$. The electron distribution enters through its cross-power spectrum with the galaxy density $\delta_g$:
$\langle\delta_g(\bk)\,\delta_e(\bk')\rangle = (2\pi)^3\delta^3(\bk\!+\!\bk')\,P_{ge}(k)$, which, on small scales, is a probe of the electron profile in halos.

The kSZ temperature is proportional to the product of the velocity and electron fields. The signal accessible to a galaxy survey is therefore a three-point correlation between two galaxy modes and the CMB temperature. This \textit{bispectra} is dominated by its squeezed configuration, in which one galaxy mode is on large scales ($k_L$) and the other two on small scales carrying the kSZ signal ($k_S\approx\ell/\chi_*$). At the squeezed limit, this bispectra factorizes as:
\begin{multline}\label{eq:bispectrum_squeezed}
\langle\delta_g(\bk_L)\,\delta_g(\bk_S)\,T(\bl)\rangle = -i\,(2\pi)^3\delta^3\!\Big(\bk_L+\bk_S+\frac{\bl}{\chi_*}\Big)\\
\times\;\frac{K_*\,k_{Lr}}{\chi_*^2}\,\frac{P_{gv}(k_L)}{k_L}\,P_{ge}(k_S)\,.
\end{multline}
Here $K_*$ is the kSZ radial kernel at $z_*$ of Eq.~(\ref{eq:K_def}), $k_{Lr}$ is the radial component of $\bk_L$, $P_{gv}(k_L)$ is the large-scale galaxy–velocity cross-spectrum, and $P_{ge}(k_S)$ is the target observable.
This factorization is the basis of kSZ tomography~\cite{Munchmeyer:2018eey,Deutsch:2017ybc,Deutsch2018a}, and it places $P_{ge}$ at $k_S=\ell/\chi_*$. The same bispectra also allow for the reconstruction of the large scale velocities and the measurement of $P_{gv}(k_L)$~\citep{Munchmeyer:2018eey,Deutsch:2017ybc,Deutsch2018a,Hotinli:2025tul, McCarthy:2024nik, Lague:2024czc,Lai:2025qdw, McCarthy:2025brx, Chaussidon:2026vmn, Cayuso:2021ljq, Contreras:2022zdz, Krywonos:2024mpb,Bloch:2024kpn, AnilKumar:2025wyt}.

\section{Pipeline Description}\label{sec:pipeline}

In this section, we introduce our kSZ tomography framework, with which we measure the small-scale galaxy–electron cross-power spectrum $P_{ge}(k)$ from the kSZ-induced cross-correlation of the CMB temperature with a galaxy-velocity-weighted momentum template. Our formalism is also similar to that of Ref.~\citep{Hotinli:2026wly}, which developed an estimator and a pipeline for measuring the moving-lens effect~\citep{1983Natur.302..315B,Hotinli:2018yyc} and reported its first detection.

\subsection{The simplified $P_{ge}$ estimator}\label{sec:pipeline_est_simplified}

We first construct a simplified version of our $P_{ge}$ estimator in a flat-sky ``snapshot'' geometry, a 3-d periodic box of side length $L$ at a fixed redshift $z_*$ with the CMB obtained by projecting onto one face of the box, and ignore the survey mask, lightcone evolution, and per-object weights. This simplified pipeline already contains the key idea: a galaxy-velocity-weighted momentum template, cross-correlated with the CMB temperature, picks out the small-scale galaxy--electron correlation in the squeezed limit.

Let $\delta_g(\bx)$ be a galaxy field with a 3-d comoving number density $n_g^{3d}$, and let $\hv_j(\bx)$ be a large-scale velocity reconstruction derived from $\delta_g$ assuming the continuity relation.
We use the standard linear reconstruction:
\begin{align}
\hv_j(\bk) &= (ik_j) \frac{U(k)}{k^2} \delta_g(\bk)\,,
    \label{eq:vhat_flat_pge}\\
      U(k) &\equiv
    \begin{cases}
    \displaystyle\frac{faH}{b_g}
      \frac{P_{gg}(k)}{P^{\mathrm{tot}}_{gg}(k)} & k < k_{\mathrm{max}}\,,
      \\[6pt]
    0 & \text{otherwise}\,,
    \end{cases}\nonumber
\end{align}
where $P^{\mathrm{tot}}_{gg}=P_{gg}+1/n_g^{3d}$ includes shot noise; the prefactor $ik_j/k^2$ is the continuity relation, $P_{gg}/P^{\mathrm{tot}}_{gg}$ is a Wiener filter that suppresses noise-dominated modes, and $b_g$ is the large-scale bias of the galaxy sample assumed in the filter. In what follows, we will apply a sharp cutoff at $k_{\rm max}=0.05\,\mathrm{Mpc}^{-1}$, which keeps the estimator in the squeezed regime, avoiding pseudo-linear scales and possible contributions from foregrounds. 

We weigh each galaxy by its reconstructed radial velocity, obtaining a 2-d momentum template:
\begin{equation}
\hpi(\btheta) = \sum_{i\in\mathrm{gal}} \hv_r(\bx_i)\,\delta^2(\btheta - \btheta_i)\,,
\label{eq:hpi_flat}
\end{equation}
where $v_r$ denotes the radial component of the velocity field. 
Constructed this way, the momentum template traces the large-scale radial velocity field inferred from the galaxy and the galaxy field on small-scales through the delta function since $\delta_g(\bx)=1/n_g^{3d}\sum_i\delta^3(\bx-\bx_i)$.

As in Ref.~\citep{Hotinli:2025tul}, we modify our estimator to mitigate the contamination to our estimators that, on average, adds a contribution that is slowly varying with redshift. We divide our galaxies into redshift bins $b$ and subtract the mean momentum template within each redshift bin $\bar{v}^b_r$. This `mean-subtracted' momentum template is:
\begin{equation}
\hpi(\btheta) = \sum_b\sum_{i\in b} [\hv_r(\bx_i)-\bar{ v}^b_r]\,\delta^2(\btheta - \btheta_i)\,.
\label{eq:hpi_flat2}
\end{equation}
The subtraction removes the modes of $\hat v_r$ that are coherent in redshift or constant across the sky (analogous to the integral constraint in galaxy clustering~\citep{deMattia:2019vdg}). In what follows, we will use $N_z=25$ equal-width redshift bins before projection and apply the same mean subtraction to both our data and the surrogate fields, which we introduce in \S\ref{sec:surrogate_normalization} without further description.

The cross-correlation between this momentum template and the CMB temperature $\langle T\,\hpi\rangle$ is then a four-point function, which reduces in the squeezed limit to a single Wick contraction, in which the velocity-velocity correlation can be written as: 
\begin{equation}
\big\langle v_r^{\rm true}(\bx)\,\hv_r(\bx) \big\rangle = \eta_r\,.
\label{eq:eta_def_flat_pge}
\end{equation}
The parameter $\eta_r$ in principle varies with $\bx$ on large scales, but for our purposes in this paper, it is a constant since we will measure the bispectrum on small scales in the squeezed limit. 

In the kSZ literature, $\eta_r$ is closely related to another quantity that is typically used to normalize the estimators or stacked profiles. This is the correlation coefficient between the true and reconstructed radial velocities $r_v\equiv\eta_r/[\sigma(v_r^{\rm true})\sigma(\hat v_r^{\rm})]$. Both $\eta_r$ and $r_v$ parameterize the efficiency of the velocity reconstruction. In our analysis, we will not use $r_v$ as our estimators are built from $\eta$, which we evaluate through our `surrogate fields' modeling framework in what follows. 

Reading off the angular cross-spectrum gives:
\begin{equation}
C_\ell^{\hpi T}\Big|_{\rm toy} = \eta_r\,n_g^{3d}\,K_*\,L\,P_{ge}(\ell/\chi_*)\,.
\label{eq:cltpi_toy}
\end{equation}
Written this way, the momentum–temperature and electron–galaxy correlations are related simply by a prefactor, which we refer to as the normalization of the estimator throughout this paper. 

This normalization is where our implementation differs from the literature. We calculate it using our surrogate-field framework, first introduced in~\citet{Hotinli:2025tul}. We turn the absolute normalization of $\langle\hpi T\rangle$ into an explicit catalog-level sum, with the survey mask, redshift evolution, per-object weights, and reconstruction efficiency computed within the pipeline, so the bandpowers of $P_{ge}(k)$ that we quote are in Mpc$^3$. The efficiency $\eta_r$ (or $r_v$) that sets this scale is elsewhere taken from simulation calibrations~\cite{AtacamaCosmologyTelescope:2020wtv, Hadzhiyska:2024qsl, RiedGuachalla:2025byu, Hadzhiyska:2025mvt, Qu:2026zyh}, or it is left free and marginalized over~\cite{Smith:2018bpn, Lai:2025qdw}; in this work, we will compute it from the randoms of the galaxy survey itself.

\subsection{The curved-sky estimator and its normalization}

Following the simplified version of our estimator in \S\ref{sec:pipeline_est_simplified}, we now calculate our curved-sky estimator and its normalization, which we use in this analysis. 

We begin by defining a weighted galaxy overdensity through the `galaxies-minus-randoms' prescription: 
\begin{equation}\label{eq:rho_g_pge}
\rho_g(\bx) = \sum_{i\in\mathrm{gal}} W_i\,\delta^3(\bx - \bx_i)
    - \frac{N_g}{N_r}\sum_{j\in\mathrm{rand}} W_j\,\delta^3(\bx - \bx_j)\,,
\end{equation}
where $N_g=\sum_i W_i$ and $N_r=\sum_j W_j$ are the weighted galaxy and random counts. 

The curved-sky momentum template is the spin-0 generalization of Eq.~(\ref{eq:hpi_flat}), following the same galaxies-minus-randoms prescription as $\rho_g$,
\begin{multline}\label{eq:hpi_def}
\hpi(\btheta) = \sum_{i\in\mathrm{gal}} W_i\,\hv_r(\bx_i)\,\delta^2(\btheta-\btheta_i)\\
    - \frac{N_g}{N_r}\sum_{j\in\mathrm{rand}} W_j\,\hv_r(\bx_j)\,\delta^2(\btheta-\btheta_j)\,,
\end{multline}
with $\hv_r$ obtained by applying the filter $U(k)$ of Eq.~(\ref{eq:vhat_flat_pge}) to $\rho_g$ at the central redshift $z_*$.
The definition of the curved-sky momentum template is similar to Eq.~\eqref{eq:hpi_flat}, but note that we include a per-galaxy weight $W_i$ that can account for completeness, FKP weighting, and other complications. As in our previous works~\citep{Hotinli:2025tul,Hotinli:2026wly}, we will choose $W_i$ to downweight galaxies with large photo-$z$ errors. 

In what follows, we compute the cross correlation of this momentum template with masked and filtered CMB temperature maps which we describe below, and compute:
\begin{equation}\label{eq:Cl_piT}
C_\ell^{\hpi T} = \frac{1}{2\ell+1}\sum_m \hpi_{\ell m}\,T_{\ell m}^{*}\,.
\end{equation}

The correlation between the temperature and momentum template is related to $P_{ge}$ with an overall normalization. In this section, we calculate this normalization and complete the definition of our curved-sky estimate.

Taking into account the survey mask, lightcone evolution, and per-object weights we defined above, the momentum template of Eq.~(\ref{eq:hpi_def}) is equal to the line-of-sight projection: 
\begin{equation}
\hpi(\btheta)=\int\chi^2 d\chi\,\bar n_{3d}(\bx)\,\hvr(\bx)\,\delta_g(\bx)\,,
\end{equation}
with $\bar n_{3d}(\bx)=\langle\sum_i W_i\,\delta^3(\bx-\bx_i)\rangle$. The observed CMB temperature is also masked and convolved with the beam in this work. The kSZ signal is given in Eq.~\eqref{eq:tksz_los}. 

The cross-correlation of the momentum template and CMB temperature traces the small-scale galaxy--electron correlation $\langle\delta_g(\bx)\, \delta_e(\bx'')\rangle_S=\zeta_{ge}(\bx-\bx'')$, and $\eta_r$. While $\eta_r$ varies slowly across the sky, the galaxy--electron correlation is local. The corresponding correlation $\langle T\hat\pi\rangle$ includes an integral of $\zeta_{ge}$, which can be written using the Limber approximation:
\begin{equation}
\int\!\!d^2\btheta''\,d\chi''\, b(\btheta'-\btheta'')\,\zeta_{ge}(\bx-\bx'') = \frac{1}{\chi^2}\,\vpot(\btheta-\btheta')\,,
\end{equation}
where $b(\btheta)$ is the CMB beam, and
\begin{equation}
\vpot(\btheta-\btheta') = \int\!\frac{d^2\bell''}{(2\pi)^2}\,b_{\ell''}\,\Pge(\ell''/\chi,\chi)\,e^{i\bell''\cdot(\btheta-\btheta')}\,.\label{eq:I_upsilon}
\end{equation}

The remaining angular integrals in $\langle T\hat\pi\rangle$ bring in the velocity-reconstruction efficiency $\eta_r$ and collapse the cross-spectrum onto a single integral over the volume:
\begin{multline}\label{eq:hpiT_eta}
\langle\hpi(\bell)\,T(\bell')\rangle = \frac{1}{4\pi}\int d^3\bx\;\bar n_{3d}(\bx)\,\frac{K(\chi)}{\chi^2}\,W_{\rm CMB}(\btheta)\\
\times\,\eta(\bx)\,b_\ell\,\Pge(\ell/\chi,\chi)\,(2\pi)^2\delta^2(\bell+\bell')\,.
\end{multline}
We approximate $\Pge(\ell/\chi,\chi)$ as equal to $\Pge(\ell/\chi_*)$ at the effective distance $\chi_*=\chi(z_*)$. Within the central two-thirds of our galaxy sample (which is inside the redshift range $0.55 < z <0.91$), $P_{ge}(k)$ (calculated from the Battaglia gas profile~\citep{Battaglia:2016xbi} and the halo model described in~Appendix~\ref{app:halo_model}) remains within $10$ percent of its values at $z_*$.  

We calculate the spatial integral in Eq.~\eqref{eq:hpiT_eta} by discretizing in:
\begin{equation}
\int d^3\bx\,\bar n_{3d}(\bx)\to\sum_{j\in\mathrm{gal}}W_j\,,
\end{equation}
which reduces our calculation to a catalog-level sum in our simulations. Eq.~\eqref{eq:hpiT_eta} recovers the boxed result $C_\ell^{\hpi T}=\N\,b_\ell\,\Pge(\ell/\chi_*)$ of Eq.~(\ref{eq:cltpi_toy}), with $\N$ being defined as a galaxy-catalog sum.

\subsection{Evaluating the normalization}\label{sec:surrogate_normalization}

In this work, we perform our analysis on a photometric galaxy survey. In order to incorporate the photometric redshift errors, we define for each galaxy the observed (photometric) and true redshifts, $z_{\rm obs}$ and $z_{\rm true}$, and the photo-$z$ errors $\sigma_z$; by deconvolving the observed $(z_{\rm obs},~\sigma_z)$ values from the galaxy catalog to infer $z_{\rm true}$, which is unknown; following Section~III.E and Appendix~B of~Ref.~\cite{Hotinli:2025tul}. We infer for both the galaxies and random samples the triple $(z_{\rm obs},\,z_{\rm true},\,\sigma_z)$ and define the comoving positions at observed and true redshifts as $\bx_{\rm obs}=(\btheta,\,z_{\rm obs})$ and  $\bx_{\rm true}=(\btheta,\,z_{\rm true})$, respectively.

We evaluate $\eta_r$ using our surrogate-field Monte-Carlo framework with the random catalogs.
Each realization draws a Gaussian density field $\delta$ from the linear power spectrum $P_{\rm lin}(k,z{=}0)$ on the analysis grid. Throughout, we fix the cosmology to \textit{Planck}~18+BAO from~Ref.~\citep{Planck:2018vyg}. From $\delta$, we form two fields: (i) the \emph{true} radial velocity through a linear continuity relation, evaluated at the random true-redshift positions and scaled by the linear growth function $D(z)$; and (ii) a \emph{surrogate galaxy field}, which is read at the true position $\bx^{\rm true}_\beta$ of each random galaxy and deposited at its observed position $\bx^{\rm obs}_\beta$:
\begin{equation}\label{eq:Sg_surrogate}
S_g(\bx) = \frac{N_g}{N_r}\!\sum_{\beta\in{\rm rands}}\,\!W_\beta\,b_g\,D(z_{\rm obs})\,\delta(\bx^{\rm true}_\beta)\delta^3(\bx-\bx_\beta^{\rm obs})\,.
\end{equation}
Applying the velocity filter $U(k)$ of Eq.~(\ref{eq:vhat_flat_pge}) to $S_g$ and evaluating at the observed-redshift positions yields the \emph{reconstructed} radial velocity $\hvr$ estimate from the surrogate fields.
With this realization, we estimate the normalization of the estimator as a sum over the random catalog:
\begin{align}\label{eq:N_surrogate}
\N &= \frac{1}{4\pi}\,\frac{N_g}{N_r}\!\sum_{\beta\in\mathrm{rand}}\!W_\beta\,\eta^{\rm surr}_r(\bx_\beta)\,W_{\rm CMB}(\btheta_\beta)\,\frac{K(\chi_\beta)}{\chi_\beta^2}\,,\nonumber\\
\eta^{\rm surr}_r&=\hvr\,v_r^{\rm true}\,,
\end{align}
where $\eta^{\rm surr}_r$ is the multiplication of the true and reconstructed velocity surrogate fields. We apply the same sharp cutoff to $\hat v_r$ at $k_{\rm max}=0.05\,\mathrm{Mpc}^{-1}$ as we do for the data to avoid pseudo-linear scales where the factorization and the linear-theory velocity weights would no longer hold, and our estimate of $\eta_r$ would become inaccurate. 

At the limit $N_r\gg N_g$ and with this sharp cutoff in $k$, the random-catalog sum gives an unbiased realization of the galaxy-catalog sum [which we show explicitly in Eq.~(\ref{eq:ClpiT_curved})]~(see a related derivation and proof in Ref.~\citep{Hotinli:2025tul}).
We find that the normalization converges after averaging over $\sim50$ independent Gaussian realizations; and use the mean of the 50 realizations to calculate the normalization $\mathcal N$ of the measured bandpowers.

\subsection{Summary of this section}
\label{ssec:curved_sky_pge}

Our pipeline reconstructs a large-scale radial velocity $\hv_r$ from the galaxy density field, weights each galaxy by it to form a momentum template $\hpi(\btheta)$, and cross-correlates $\hpi$ with the CMB temperature; in the squeezed limit, that cross-correlation returns the small-scale galaxy–electron power spectra.

We followed the curved-sky pipeline of Ref.~\citep{Hotinli:2025tul} and introduced in this section the velocity filter, the momentum template of Eq.~(\ref{eq:hpi_def}), and the cross-spectrum of Eq.~(\ref{eq:Cl_piT}). We have shown that the curved-sky cross-spectrum is:
\begin{equation}\label{eq:ClpiT_curved}
\boxed{\;
\begin{aligned}
C_\ell^{\hpi T} &= \N\,b_\ell\,P_{ge}(\ell/\chi_*)\,,\\
\N &\equiv \frac{1}{4\pi}
\sum_{j\in\mathrm{gal}} W_j\,\eta_r(\bx_j)\,W_{\rm CMB}(\btheta_j)\,
\frac{K(\chi_j)}{\chi_j^2}\,.
\end{aligned}\;}
\end{equation}
Here, $W_{\rm CMB}$ is the CMB mask, $K(\chi)$ the kSZ kernel along the lightcone, $b_\ell$ is the CMB beam, and $\eta_r(\bx)\equiv\langle\hv_r(\bx)\,v_r^{\rm true}(\bx)\rangle$ is the reconstruction efficiency, averaged over realizations of the matter field at fixed survey geometry.
In our formalism, $C_\ell^{\hpi T}$ is a pseudo-power spectrum, with no $f_{\rm sky}$ correction; where the partial-sky and mask effects are absorbed into $\N$.
The two sides treat redshift differently: $\Pge$ is a single small-scale spectrum read at $z_*$, while $\N$ sums over the galaxies at their own redshifts.
$\N$ puts the measured bandpowers in physical units; the significances, the null tests, and the $P_{ge}$ template shape in $\ell$ do not depend on it.

\section{Data}\label{sec:data}

\begin{figure*}[t]
\centering
\includegraphics[width=\textwidth]{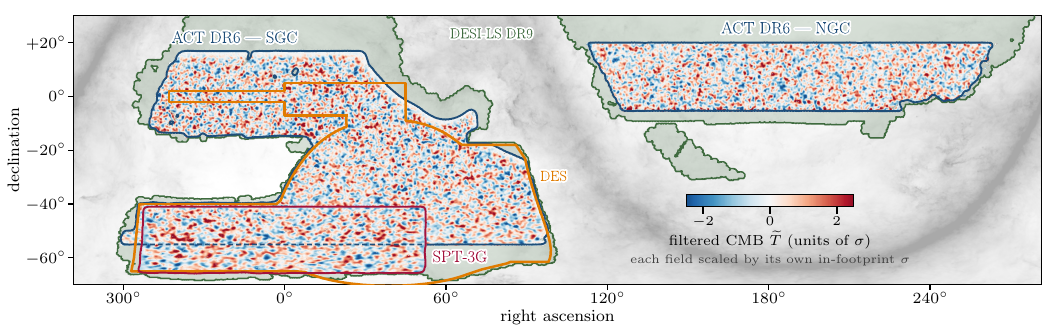}
\vspace*{-24pt}
\caption{The filtered CMB temperature, filtered for display over the three analysis footprints, in equatorial coordinates and in units of the in-footprint standard deviation of each field. The two ACT~DR6 caps are disjoint; the SPT-3G field lies inside the southern cap, and the dashed blue line marks the southern edge of the ACT selection.
Where the two overlap the SPT-3G field is drawn on top. Two surveys are drawn for orientation: the DESI-LS DR9 footprint~\cite{Dey:2018}, from which both galaxy selections are cut, in green, and the DES footprint~\cite{DES:2016jjg} in orange. We filter the CMB multipoles with $\propto b_\ell\,P_{ge}/P_{gg}/C_\ell^{\rm tot}$ on $\ell\in[2000,6000]$ for the figure only. Planck 353\,GHz dust~\cite{Planck:2016frx} is grayscale outside the footprints.}
\label{fig:footprint}
\end{figure*}

We now describe the CMB maps and the galaxy catalogs we use in this analysis. We use CMB temperature maps from ACT~DR6~\cite{Naess2025,ACT:2023wcq} and SPT-3G~D1~\cite{SPT-3G:2026krg}, and for the galaxy samples, we use the ``main'' and ``extended'' photometric LRG selections of the DESI Legacy Imaging Surveys DR9~\cite{Dey:2018,DESI:2022gle,Zhou:2023gji}.

\subsection{CMB maps}\label{sec:data_cmb}

We use the ACT~DR6 single-frequency source-free temperature coadds at 98, 150, and 220\,GHz~\cite{Naess2025}; we call the 98\,GHz channel ``90\,GHz'' throughout, as the ACT data products do. The coadds come in two versions: day$+$night, combining daytime and nighttime observations, and night, from nighttime data alone. The day$+$night maps are deeper; while at the same time, the daytime beam is less well characterized because the sun heats the primary mirror~\cite{Naess2025}.

We adopt the night maps for our baseline results, a choice we make for the robustness of our results, which also matches the night-only construction of the DR6 NILC map below and of other DR6 kSZ analyses~\cite{ACT:2023wcq, Hadzhiyska:2024qsl, RiedGuachalla:2025byu, Qu:2026zyh, Hadzhiyska:2026uyq}, and quote the day$+$night results alongside throughout Section~\ref{sec:results}. We use the public ACT~DR6 needlet internal linear combination (NILC) map~\cite{ACT:2023wcq} as a foreground-cleaned channel, which preserves the blackbody CMB$+$kSZ signal at a common $1.6'$ FWHM beam.

For ACT, we use the Planck $f_{\rm sky}=0.7$ Galactic mask together with a pixel-weight function that keeps the pixels where the ACT map noise is below $70\,\mu\mathrm{K}$-arcmin at both $90$ and $150$\,GHz following~\cite{Hotinli:2025tul}, and below $150\,\mu\mathrm{K}$-arcmin at $220$\,GHz. The night-map depths are $15$, $16$ and $74\,\mu\mathrm{K}$-arcmin at $90$, $150$ and $220$\,GHz, median over the exposed area~\cite{Naess2025}. 

We make a few analysis choices based on null tests. In the ACT Southern Galactic Cap (SGC), we add the ACT~DR6 lensing-analysis mask~\cite{ACT:2023kun,ACT:2023dou}, and on the day$+$night maps, we drop the declination band $-20^\circ<\delta<0^\circ$, where the Legacy Surveys imaging changes hands from deeper Dark Energy Survey data to DECaLS-only imaging~\cite{Dey:2018,Zhou:2023gji}.
The lensing mask removes $9\%$ of the southern galaxies, and the declination band removes a further $30\%$, so the southern night and day$+$night numbers rest on different sky selections, each with its own $\N$; Section~\ref{ssec:nulls} gives their effect on the amplitudes and on the null tests.

For the SPT-3G maps we use the full-depth SPT-3G~\cite{SPT-3G:2014dbx,SPT-3G:2021vps}~D1 temperature coadds at 95, 150, and 220\,GHz~\cite{SPT-3G:2026krg}, formed from the first two years (2019--2020) of observations of the SPT-3G Main field; the SPT$\times$DESILS analysis footprint is the overlap of the field with the DESILS-LRG sample, $-66^\circ\lesssim\delta\lesssim-41^\circ$ (Section~\ref{sec:data_gal}). We apply the apodized mask released with the SPT-3G maps~\cite{SPT-3G:2026krg}, which discard the noisy field edges, keeping sky where all three channels are well measured.

The released weight maps from SPT-3G D1 are normalized to their peak, and we keep pixels where the weight of a channel exceeds $5\%$ of it ($10\%$ at $220$\,GHz), intersected over the three channels so that all frequencies are analyzed on the same sky and the frequency differences are well defined; the released apodized mask tapers the survey borders and excises high-significance sources and galaxy clusters. We process the SPT-3G maps at $N_{\rm side}=4096$, sufficient for the $\ell_{\rm max}=9000$ to which we compute spectra.

We use the SPT-3G effective beam for $b_\ell$ of Eq.~(\ref{eq:ClpiT_curved}) made available from the SPT-3G D1 public data products. The SPT effective beam multiplies three terms: the SPT-3G CMB-SED beam from Chaubal~et~al.~\cite{SPT-3G:2026mcr}, the filter transfer function released with these maps and the SPT-3G Main-field pixel window (within 1\% of the full-sky HEALPix pixel window at $\ell=10^4$).
All three enter multiplicatively, folded into the common transfer function $b_\ell$ of Eq.~(\ref{eq:ClpiT_curved}) rather than deconvolved from the maps, so that the SPT temperature is cross-correlated with $\hpi$ without map-level filtering beyond the mask and the beam-equalization which we describe next. 

Last, throughout this work, we beam-equalize the higher-frequency channels of both ACT and SPT-3G maps to their widest beams—that is, the lowest frequency (90\,GHz for ACT, 95\,GHz for SPT-3G)—in harmonic space, such that we filter the CMB multipoles by $b_\ell^{\nu}$ over $b_{\ell}^{90}$ (or $b_{\ell}^{95}$) for $\nu=\{150,220\}$. This lets the $P_{ge}$ measurements share a common beam $b_\ell$ separately for each survey in our analysis.

\subsection{Galaxy catalogs}\label{sec:data_gal}

We use the ``main'' LRG sample of~Ref.~\cite{Zhou:2023gji}, identical to the DESI LRG target selection~\cite{DESI:2022gle}, and the ``extended'' sample of the same reference, whose looser color and magnitude cuts give a higher number density at the cost of larger photometric-redshift errors; both are selected from the DESI Legacy Imaging Surveys DR9 imaging~\cite{Dey:2018} and share its footprint.
We restrict ourselves to photometric redshifts $0.4\le z_{\rm obs}\le 1.1$, apply the veto masks and quality cuts of~Refs.~\cite{White:2021yvw,Zhou:2023gji}, and use only the galaxies and randoms on the sky within our CMB masks.

After the redshift, quality, and $W_{\rm CMB}>0$ cuts, the galaxy counts are those of Table~\ref{tab:footprints}, at comoving densities $\bar n_g(z_*)=1.50\times10^{-4}$ (main) and $3.46\times10^{-4}\,\mathrm{Mpc}^{-3}$ (extended). The ACT analysis footprint ends at $\delta=-55^\circ$; the $973\,\mathrm{deg}^2$ of the SPT-3G footprint north of that edge, and $64\%$ of its galaxies (main and extended alike), also enter the ACT SGC analysis.

After all cuts, we use about $220$ randoms per galaxy in each ACT cap for the main sample, about $85$ for the extended sample, and about $115$ and $45$ in the SPT-3G main and extended samples.
Figure~\ref{fig:footprint} shows the resulting footprints.

\begin{table}[t]
\centering
\caption{Analysis footprints and galaxy counts.  Areas are approximate; the day$+$night southern selection is the night sky less the excised declination band (Section~\ref{sec:data_cmb}).}
\vspace{1.0mm}
\small
\begin{tabular}{@{}l @{\hspace{2.5mm}} r @{\hspace{3mm}} r @{\hspace{2mm}} r@{}}
\toprule
 & & \multicolumn{2}{c}{$N_g$} \\
\cmidrule(l){3-4}
Footprint & Area [deg$^2$] & main & extended \\
\midrule
ACT NGC                    & $3310$ & $1{,}670{,}020$ & $4{,}342{,}017$ \\
ACT SGC, night             & $6180$ & $3{,}110{,}183$ & $8{,}063{,}422$ \\
ACT SGC, day$+$night       & ---    & $2{,}182{,}358$ & $5{,}649{,}798$ \\
SPT-3G$\times$DESILS   & $1520$ & $761{,}864$     & $1{,}975{,}785$ \\
\bottomrule
\end{tabular}
\label{tab:footprints}
\end{table}

We test two definitions for our per-object weights $W_i$. First, we define the weights as:
\begin{equation}\label{eq:weight_def}
W_i = \exp\!\left[-\frac{\sigma_{z,i}^2}{\alpha\,(1+z_i)^2}\right]\,,
\hspace{1cm}
\alpha = 2.5\times 10^{-3}\,,
\end{equation}
with $\sigma_{z,i}=\sigma_z(1+z_i)$, which down-weights the galaxies that are assigned large photo-$z$ errors in the DESILS-LRG catalog compared to the correlation length of the velocity field. Next, we adopt a single photometric scatter per sample, $\sigma_z=0.024$ (main) and $0.027$ (extended), from the spectroscopic calibration of Ref.~\cite{Zhou:2023gji}. We find our results are insensitive to our choice on this and adopt constant errors from Ref.~\cite{Zhou:2023gji} throughout. As in Ref.~\cite{Hotinli:2025tul}, we apply no imaging or FKP weights.

\section{Results}\label{sec:results}

Having developed the formalism for our $P_{ge}$ measurement and described the data we use, we now describe our analysis and results in this section. Our main result is that our measurements of $P_{ge}$ agree between the ACT and SPT maps. We first describe our statistical analysis choices in \S\ref{sec:statmethod_pge} and present our results in the rest of this section.  

\subsection{Statistical analysis}\label{sec:statmethod_pge}

We measure $C_\ell^{\hpi T}$ using Eq.~\eqref{eq:ClpiT_curved} and bin this cross-spectrum into $N_b=13$ bandpowers of equal width in $\ell$ over $\ell\in[1500,7500]$, which, at the effective comoving distance of the sample $\chi_*\simeq2679$\,Mpc, correspond to wavenumbers $k\in[0.56,2.80]\,\mathrm{Mpc}^{-1}$.

We define $D_\ell\equiv \ell\,C_\ell^{\hpi T}$ and $w_\ell\equiv (2\ell+1)$, and write the binned data vector and the normalized bandpowers as:
\begin{equation}\label{eq:l_binning_pge}
d_b = \frac{1}{W_b}\sum_{\ell\in b} w_\ell\,D_\ell\,,
\hspace{0.7cm}
P_{ge,b} \equiv \frac{1}{\N\,W_b}\sum_{\ell\in b} w_\ell\,C_\ell^{\hpi T}\,,
\end{equation}
with $W_b \equiv \sum_{\ell\in b} w_\ell$. The second equality measures the galaxy-electron correlation multiplied by the CMB beam, $b_\ell\,P_{ge}(k)$ in Mpc$^3$. We do not deconvolve our estimator with the CMB beams in our analysis, since the same $b_\ell$ will multiply our theoretical prediction of $P_{ge}$ in our comparison below, and it cancels in the fitted amplitude. Nevertheless, the figures divide the beams out for display.

We calculate the bandpower covariance $\mathbf{C}$, with entries $C_{bb'}=\Cov(d_b,d_{b'})$ in two different ways. First, we build the covariance using the data alone. We define:
\begin{equation}\label{eq:cov_data}
C_{bb'} = \frac{1}{W_b W_{b'}}\!\!\sum_{\substack{\ell\in b,\ \ell'\in b'\\ |\ell-\ell'|\le\Delta\ell}}\!\! w_\ell\,w_{\ell'}\,(D_\ell-d_b)(D_{\ell'}-d_{b'})\,,
\end{equation}
summed over multipole pairs within a window $\Delta\ell$ set to the mode-coupling scale of each footprint (we set $\Delta\ell=11$ and $5$ for the ACT NGC and SGC caps, and $10$ for SPT-3G). Because $\Delta\ell$ is much smaller than the width of a bandpower, only adjacent bins share coupled multipole pairs; therefore, the covariance is essentially tridiagonal, and bins that are two or more apart are uncorrelated.
We find, moreover, that the data show nearest-neighbor correlations of at most seven percent in every dataset and channel.

Second, we use the scatter of signal-free surrogate realizations of the momentum template we defined in \S\ref{sec:surrogate_normalization}, which we cross-correlate with the measured (not simulated) CMB maps, where we use $300$ realizations per ACT northern cap, $200$ per southern cap, and $500$ per SPT-3G sample to estimate the covariance. We calibrate this surrogate template power to that of the data by one factor per dataset to account for the clustering of the data. We find the ratio between the data covariance and the surrogate is flat in $\ell$ to $2$--$3\%$.

We find the median difference between the two methods to be within $5\%$, over the thirteen bins of every channel and dataset. The two depart most in the lowest bin and between $\ell\simeq3300$ and $4300$. Throughout this paper, the uncertainties, results, and null-test values quoted come from the surrogate ensemble calculation of the covariance. We use diagonal entries of our covariance components in the 13 $\ell$-bins, as the bin-to-bin correlations are small, and do not enter our analysis.

We summarize each measurement by a single amplitude $\bksz$, the ratio between the measured bandpowers and our halo-model prediction of $P_{ge}$. We calculate $P_{ge}$ from a halo occupation distribution matched to the measured number density of each sample and from electrons following the Battaglia gas profile~\cite{Battaglia:2016xbi}. 

For our main results, we truncate the gas profiles at the radius where the total enclosed gas mass reaches the cosmic baryon budget of the halo (i.e. total enclosed halo mass times the baryon fraction), so that $\bksz=1$ corresponds to gas distributed according to this truncated profile. These radii are typically around 2-3 times the virial radius of the halos in the mass range of LRGs for the Battaglia profile. 

We also fit the amplitude for a profile we calculate by truncating it at the virial radius of halos and normalizing it so that the enclosed gas mass is equal to the halos' virial mass times the gas fraction, as was used in other works~\citep[e.g.][]{Hotinli:2026wly,Chaussidon:2026vmn}. Truncating the profile at the virial radius gives a second template rather than a rescaling of the first, since both are normalized to the mass they enclose, and the second holds the same gas mass in a more concentrated way.

The single amplitude describes the bandpowers well, at $\chi^2$ probabilities between $0.54$ and $0.81$ for our main results. We also explored allowing the shape to vary as well, with four free parameters in place of one; this does not improve the fit and leaves the enclosed gas fraction unchanged, as we show below. We therefore use the scaled template as an interpolation of $\Pge$ across the measured range rather than as a statement about the shape of the gas profile, and we give its construction in Appendix~\ref{app:halo_model}.

We fit the data vector $\boldsymbol{d}=\{d_b\}$ to this single-amplitude template $\boldsymbol{t}=\{t_b\}$ by minimizing: 
\begin{equation}
    \chi^2(\bksz)=(\boldsymbol{d}-\bksz\,\boldsymbol{t})^T\mathbf{C}^{-1}(\boldsymbol{d}-\bksz\,\boldsymbol{t})\,,
\end{equation}
giving
\begin{equation}\label{eq:bksz_fit_pge}
\begin{gathered}
\hbksz = \frac{\boldsymbol{t}^T\,\mathbf{C}^{-1}\,\boldsymbol{d}}{\boldsymbol{t}^T\,\mathbf{C}^{-1}\,\boldsymbol{t}}\,,
\hspace{1cm}
\sigma^2(\bksz) = \frac{1}{\boldsymbol{t}^T\,\mathbf{C}^{-1}\,\boldsymbol{t}}\,,
\end{gathered}
\end{equation}
and
\begin{equation}
\frac{\hbksz}{\sigma(\bksz)} = \frac{\boldsymbol{t}^T\,\mathbf{C}^{-1}\,\boldsymbol{d}}{\sqrt{\boldsymbol{t}^T\,\mathbf{C}^{-1}\,\boldsymbol{t}}}\,.
\end{equation}

We treat the two caps (NGC and SGC) as statistically independent, stack their normalized bandpowers, each formed with its own $\N$, into a single $2N_b$-vector with a block-diagonal covariance, and fit the same template in Mpc$^3$; combining in Mpc$^3$ rather than in $C_\ell^{\hpi T}$.
The caps share a parent catalog, selection, and redshift range, so we expect a common $P_{ge}(k)$, which the cap-difference null of Section~\ref{ssec:nulls} tests. We report the goodness of fit as the reduced $\chi^2$ over the number of degrees of freedom $N_{\rm dof}$, and the probability-to-exceed ${\rm PTE}_{\rm gof} = \Pr(\chi^2 > \chi^2_{\rm obs})$, and consider a fit to pass if ${\rm PTE}_{\rm gof}$ exceeds 0.05. 

The estimator is linear in the CMB temperature, so any component uncorrelated with the velocity-weighted template, such as the primary CMB, lensing, and instrument noise, adds variance but no bias~\cite{Smith:2018bpn}.
The residual concern is the extragalactic foregrounds; thermal Sunyaev Zel'dovich (tSZ) effect, the cosmic infrared background (CIB) and radio sources; which trace the same structure as the galaxies, while the kSZ signal has a blackbody spectrum~\cite{EmbilVillagra:2026fnh}. In this paper we bound them from the data with the null tests of Section~\ref{ssec:nulls}.

\subsection{$P_{ge}$ signal measurements}\label{ssec:signal}

\begin{figure*}[t!]
\centering
\includegraphics[width=\textwidth]{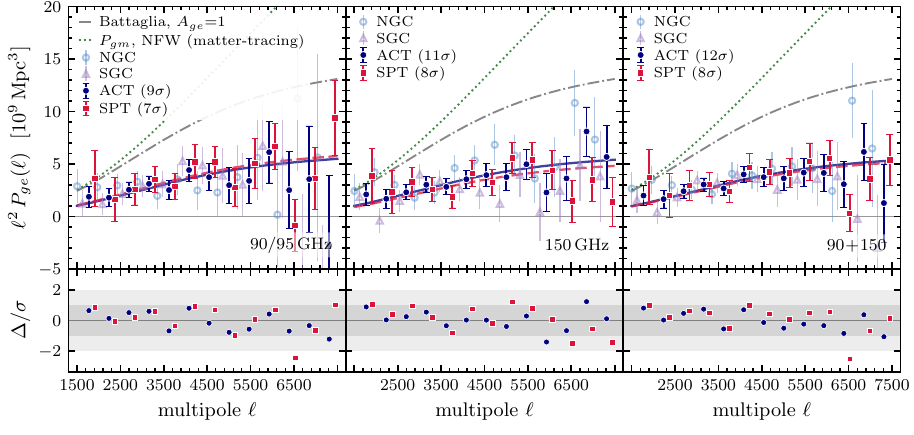}
\vspace*{-24pt}
\caption{Measured $P_{ge}$ bandpowers for the DESILS-LRG main sample on the night maps, at $\sim$\,$90/95$\,GHz (left), $150$\,GHz (centre), and the $90+150$\,GHz combination (right), in $13$ bins of equal width in $\ell\in[1500,7500]$ (Table~\ref{tab:detections}).
Navy circles: the ACT DR6 inverse-variance NGC$+$SGC combination (faint open markers: the two caps separately); red squares: SPT-3G.
The $90\!+\!150$ combination is a per-footprint average of the two beam-equalized frequency spectra, $(C^{90}_\ell+C^{150}_\ell)/2$; and is listed in Table~\ref{tab:detections}.
The navy solid and red dashed curves are the Battaglia template~\citep{Battaglia:2016xbi} rescaled to the amplitude fitted to each survey, over the fitted range only.
Grey dash-dotted: the full-budget Battaglia template ($\bksz=1$); green dotted: a matter-tracing NFW $P_{gm}$ for reference.
Bandpowers are shown with the beam divided out (Section~\ref{sec:pipeline}).
Lower subpanels are per-bin residuals from the best fit of each survey in units of $\sigma$, with $\pm1$ and $\pm2\sigma$ bands.
}
\label{fig:bandpowers}
\end{figure*}

We now describe our main results in this paper: our measurement of the $P_{ge}$ bandpowers with SPT-3G and ACT maps. For each survey and cap, we estimate the cross-spectrum $C_\ell^{\hpi T}$ as in~Eq.~\eqref{eq:ClpiT_curved}, bin it $N_b=13$ bandpowers of Eq.~(\ref{eq:l_binning_pge}), calculate its covariance using our surrogate-field framework in~\S\ref{sec:surrogate_normalization}, and fit the single amplitude $\bksz$ of Eq.~(\ref{eq:bksz_fit_pge}).
Throughout, we quote the night coadds and share the day$+$night results alongside in Tables~\ref{tab:detections}-\ref{tab:nulls}.

\subsubsection{Measurements of the $P_{ge}$ amplitude with SPT and ACT}

Figure~\ref{fig:bandpowers} shows the bandpowers for frequencies 90/95, 150, and their combination for each survey and footprint. The two caps agree, with every channel on the night coadds within $2.5\sigma$ and every channel but $150$\,GHz within $1.7\sigma$. The Table~\ref{tab:detections} collects the per-channel amplitudes with their significances $\bksz/\sigma(\bksz)$, which range from $5.8$ to $11.8$ across the tabulated night channels.

The two surveys measure the same amplitude at every channel.
The ACT two-cap and SPT-3G amplitudes are $0.419\pm0.048$ against $0.439\pm0.059$ at $90/95$\,GHz, $0.413\pm0.038$ against $0.368\pm0.047$ at $150$\,GHz, $0.408\pm0.035$ against $0.392\pm0.047$ in the $90+150$ combination, and $0.342\pm0.136$ against $0.409\pm0.128$ at $220$\,GHz, every channel agreeing within $0.8\sigma$. {For either main or extended samples, the $90+150$ difference between the two surveys remains within $-0.03$ and $+0.31\sigma$.}

Changing our analysis choices, such as the southern sky selection or changing the ACT night maps to day+night, does not affect the amplitude by more than $0.8\sigma$, and we fixed both choices before computing the amplitudes.
The normalization of the template enters $\bksz$ and $\sigma(\bksz)$ in the same way and cancels in the significance, so across the two occupations we considered based on the truncation radius of the gas profiles, every significance in Table~\ref{tab:detections} moves by less than $1.5\%$.

The $220$\,GHz maps are much noisier than those at $150$\,GHz and add little to the detection, so we leave that channel out of Table~\ref{tab:detections} and use it in the null tests, where a frequency-dependent contaminant would show up most strongly.
In \S\ref{ssec:nulls} we find that all \Ntwotwentymainw{} $220$\,GHz frequency differences are within $0.7\sigma$ of zero.

The agreement is closest in the southern cap, where the two surveys observe the same galaxies and the main-sample $150$\,GHz amplitudes are $0.318\pm0.055$ (ACT) and $0.368\pm0.047$ (SPT-3G).
In all differences in $P_{ge}$ between SPT and the two caps of ACT  pairs sharing the same galaxy sample and frequency channel, none exceed $1.9\sigma$. 
The difference that reaches it is SPT vs ACT NGC, which shares no sky.
Combining SPT-3G and ACT NGC gives $\bksz=0.416\pm0.034$ and the significance reaches $12.3\sigma$ on the main sample, against $11.7\sigma$ for the two ACT caps together (Table~\ref{tab:detections}); the two independent footprints fit together at a $\chi^2$ probability of $0.76$.

Across both surveys and every channel, the fitted amplitude lies between $\bksz=0.318$ and $0.507$ (Table~\ref{tab:detections}) for the night maps, which is a factor of $2.0$ to $3.1$ below the $A_{ge}=1$ prediction, and $2.5$ for the ACT $90+150$ combination. 
Fitting the same bandpowers to the virial-radius template lowers the amplitudes to $0.68$--$0.77$ of the values in Table~\ref{tab:detections}; a factor of $3.3$ below $A_{ge}=1$ for this template, and moves the detection significances by at most $0.27\sigma$ (Table~\ref{tab:detections_rvir}).
The virial-radius template describes the bandpowers less well, with two SPT-3G entries below a $\chi^2$ probability of $0.05$ against none for our main choice of truncation.

We quote the fixed-profile fit throughout.
A gNFW shape with four free parameters leaves the gas fraction unchanged within $1\sigma$ and fits the bandpowers no better ($\Delta\chi^2$ between $+0.7$ and $+2.4$ for three extra parameters, against $\simeq3$ expected by chance).
Our measurement, therefore, constrains the amount of gas rather than its arrangement, and we provide the calculation in Appendix~\ref{app:halo_model}.

\subsubsection{Measuring the gas fraction}

Turning the fitted template into a gas fraction requires the mean mass of the halos that host the galaxies, which the cross-spectrum alone does not measure. Hadzhiyska, Ferraro \emph{et al.} fit a halo occupation model to the ACT~DR6 CMB lensing of the DESILS-LRG galaxies, which constrains the halo mass directly, with a systematic uncertainty of about $7\%$ on the mean halo mass~\cite{Hadzhiyska:2025mvt}. Evaluated with the mass function of our halo model, that occupation gives $\log_{10}(M_{\rm vir}/h^{-1}M_\odot)=13.23$.

Adopting that occupation as published and fitting our own bandpowers, the two ACT caps combined place $32.4\pm2.8\%$ of the cosmic baryon budget of the halo within the virial radius, SPT-3G places $31.0\pm3.7\%$, and SPT-3G and ACT NGC combined, $33.0\pm2.7\%$. The Battaglia profile with $A_{ge}=1$ places $62\%$ of the budget inside the same radius. 

Our calculation of the gas fraction is also sensitive to our assumptions about the halo occupation distribution (HOD). The galaxy number density we measure is lower than what the published HOD predicts, and attributing that difference to a higher mass threshold raises the mean host mass to $13.35$, lowering the ACT measurement of the gas fraction to $25.8\pm2.2\%$, the SPT-3G to $24.8\pm2.9\%$, and the combination to $26.3\pm2.1\%$. 

The two surveys agree under both halo occupations within $0.3\sigma$, and between our two choices, the gas fraction varies by about six percent, which is approximately six times the difference between the surveys. Therefore, what limits how well the gas fraction can be inferred is the occupation and not the measurement.
Between them, the two occupations and the three combinations bracket the gas fraction at $25$--$33\%$.

\begin{figure*}[t!]
\centering
\includegraphics[width=\textwidth]{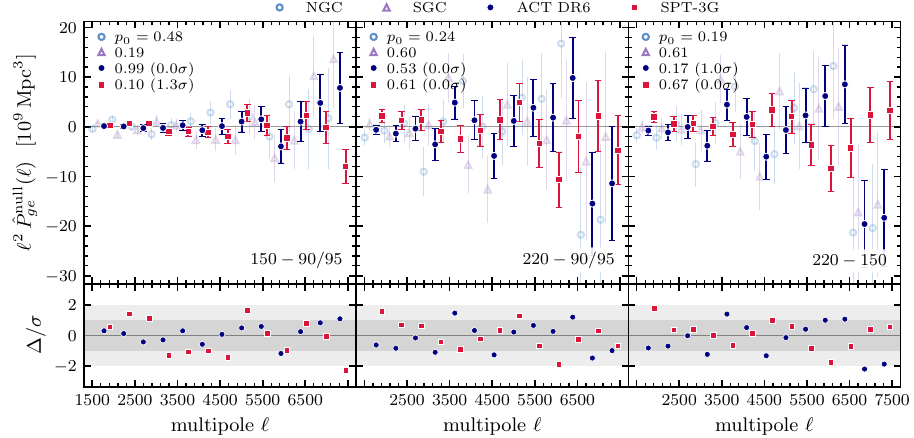}
\vspace*{-21pt}
\caption{Frequency-difference null bandpowers for the DESILS main sample on the night maps at their corresponding footprint, in the binning, series and marker conventions of Figure~\ref{fig:bandpowers}. The columns are the $150-90/95$, $220-90/95$, and $220-150$ differences.
Legends quote $p_0$, and the excess in $\sigma$ it corresponds to, for SPT-3G and for the inverse-variance combination of the two ACT caps.
Every $p_0$ is computed on the beam-convolved bandpowers, not on the plotted points. 
The full null suite, including the per-cap and cap-difference nulls, is collected in Table~\ref{tab:nulls}.}
\label{fig:freqnull}
\end{figure*}

\subsection{Null tests}\label{ssec:nulls}

The kSZ signal has a blackbody spectrum, so a difference of our beam-equalized single-frequency $P_{ge}$ spectra should contain no kSZ signal, while a contaminant from foregrounds with any other frequency dependence (tSZ, CIB, radio) would survive. 

To test our measurements against foreground contamination, we form three pairwise differences of each survey (Figure~\ref{fig:freqnull}).
The two ACT caps are disjoint and are observed with the same instrument, so their difference also tests for direction-dependent systematics and for the common $\Pge$ assumed when the caps are combined; covariance is the sum of the two cap covariances.

We quantify each null test by the probability $p_0$ of obtaining a larger $\chi^2$ than zero by chance. Throughout, we count $p_0<0.05$ as a failure. This $p_0$ is distinct from the goodness-of-fit probability of the signal fits.
All $\NnullMainNight$ main-selection tests of Table~\ref{tab:nulls} pass on the night maps, the smallest being the $150$\,GHz cap difference at $p_0=0.057$ and the next $0.10$, and all $\NnullMainDn$ pass on the day$+$night maps.
Overall, Table~\ref{tab:nulls} includes $\NnullAll$ tests: 6 on SPT maps and 20 on the ACT night and day$+$night maps each.

All $\NnullExtNight$ extended tests also pass on the night maps, the smallest being at $p_0=0.06$. On the day$+$night maps \NpassExtDnw{} of the \NnullExtDnw{} extended tests pass. The only exception, and the only probability below $0.05$ anywhere in the $\NnullAll$ tests of Table~\ref{tab:nulls}, is the extended northern $220-150$\,GHz difference at $p_0=0.03$, which gives $0.06$ on the night maps. We show the frequency-difference null bandpowers for the extended sample in Figure~5.

\section{Discussion and Conclusions}\label{sec:discussion}

We have measured the galaxy–electron cross-power spectrum on small scales ($k\gtrsim0.4\,\mathrm{Mpc}^{-1}$) by cross-correlating a velocity-weighted galaxy momentum template from DESILS-LRGs with the CMB temperature from ACT DR6 and, for the first time, SPT-3G.

ACT~DR6 and SPT-3G observe galaxies drawn from the same DESILS-LRG catalogs with independent optics, beams, filtering, and noise, and return the same $\Pge$ amplitude at every channel, within $0.8\sigma$.
The largest difference of the \Npairsw{} cap-by-cap pairs, $1.9\sigma$, is against the ACT northern cap, which shares neither sky nor galaxies with SPT-3G.
Each survey detects $\Pge$ on its own, at $11.7\sigma$ and $8.4\sigma$, respectively, and the two non-overlapping footprints together, SPT-3G and ACT NGC, at $12.3\sigma$.

Instrument-specific systematics are not shared between two telescopes, so the agreement establishes SPT-3G alongside ACT for the measurement of $P_{ge}$; and the gas distribution that ACT analyses report~\cite{Hadzhiyska:2024qsl, RiedGuachalla:2025byu, Hadzhiyska:2025mvt, Qu:2026zyh, Hadzhiyska:2026uyq} no longer rests on a single instrument.
SPT-3G reaches the southern sky that ACT does not observe, and its maps are deeper; therefore, the two surveys are complementary rather than redundant.

The amplitude $A_{ge}$ that we measure lies between $0.29$ and $0.51$ in every dataset (Table~\ref{tab:detections}), consistent across both surveys and all frequencies, and every null test on SPT-3G and the ACT night maps is consistent with zero.
At this measurement precision, different definitions of the gas profile scale the predicted $\Pge$ up or down, so the amplitude and its error move together, and the significances change by less than $1.5\%$.

Read as a gas fraction, the amplitude gives $25$--$33\%$ of the cosmic baryon budget of the halo within the virial radius (Section~\ref{ssec:signal}), against $62\%$ for the same profile at $\bksz=1$.
The width of that interval is set by the assumed host halo mass rather than by the errors on the bandpowers, and its two ends correspond to calibrating the halos' galaxy occupation to the measured number density of the sample, and their masses to the CMB lensing of the same galaxies.
On any of those assumptions, a substantial part of the gas budget is not within the virial radius, which is the same picture that kSZ and lensing measurements of comparable LRG samples describe~\cite{Hadzhiyska:2024qsl,RiedGuachalla:2025byu,Qu:2026zyh,Lai:2025qdw,Hadzhiyska:2025mvt}, and the values we obtain sit in the range they report.
We do not infer where the rest of it lies, though extending the fit to lower multipoles (larger scales) would bear on that.

The measurement is already precise enough that what limits the interpretation of the amplitude is the template and the galaxy sample that define it, not the error bars.
The kSZ signal of LRGs has already been resolved in four redshift bins and in bins of stellar mass, on the photometric and on the DESI spectroscopic selections with ACT~\cite{Hadzhiyska:2024qsl,RiedGuachalla:2025byu,Qu:2026zyh}.
The halo mass that anchors our gas fraction is likewise measured rather than assumed, from the CMB lensing of the DESILS-LRG galaxies~\cite{Hadzhiyska:2025mvt}, and it supplies one end of the interval we quote.

Lower-noise maps from the Simons Observatory and the wider SPT-3G field will tighten the per-channel errors, but not the assumptions that set the width of that interval.
Analyzed alongside weak lensing, the kSZ signal measures the matter--electron cross-spectrum directly~\cite{Hadzhiyska:2026aeo,Ganguly:2026yos}, so that the baryonic suppression that upcoming lensing surveys must otherwise marginalize over is measured instead.
We take up such a joint analysis, and the redshift dependence of $\Pge$, in forthcoming work.

In the paper, we developed a method to measure the electron-galaxy power spectrum by cross-correlating the CMB and galaxies. Our method is complementary to stacking and other Fourier space analyzes in the literature, while differing in the calculation of the normalization of our estimator, which we do directly from the DESILS-LRG random catalogs via the surrogate-fields formalism we describe. We measured the galaxy-galaxy-kSZ bispectrum for the first time with the SPT-3G maps and showed that the two surveys (ACT and SPT) are in excellent agreement.

\section*{Acknowledgments}

SCH acknowledges support from grant NSF PHY-2309135 to the Kavli Institute for Theoretical Physics (KITP), where this work was initiated.
SCH also acknowledges the Aspen Center for Physics, where this work was further developed and which is supported by National Science Foundation grant PHY-2210452. We thank Kendrick Smith for contributions to the development of the method and the estimator used in this work, as well as for useful discussions; Gil Holder for directing us to the SPT-3G~D1 data products, which proved to be excellent and straightforward to work with, and Matthew Johnson and Gil Holder for comments on the manuscript; Mathew Madhavacheril and Alex Lagu\"{e} for collaboration at the initial stages of this work; and Lindsey Bleem, Edmond Chaussidon, Simone Ferraro, Boryana Hadzhiyska and Noah Sailer for useful discussions. Research at Perimeter Institute is supported in part by the Government of Canada through the Department of Innovation, Science and Economic Development Canada and by the Province of Ontario through the Ministry of Colleges and Universities.
The computations reported in this work were run on the high-performance computing cluster at Perimeter Institute.
We thank the ACT and the SPT-3G collaborations for making their maps public, and the DESI Legacy Imaging Surveys team for the galaxy catalogs.

\section*{Data and code availability}

This analysis uses public data: the ACT~DR6 single-frequency coadds~\cite{Naess2025} and the ACT~DR6 NILC map~\cite{ACT:2023wcq}, the SPT-3G~D1 maps~\cite{SPT-3G:2026krg}, and the DESI Legacy Imaging Surveys DR9 LRG catalogs~\cite{Dey:2018,Zhou:2023gji}.
The modified version of the {\tt kszx} package used in this work, together with the analysis code that produces the numbers and figures of this paper, will be made public upon publication and is available from the author on request before then; the measured bandpowers and their covariances are available from the author on request.

\appendix

\section*{Appendices}

In the appendices below, we evaluate $r_v$ and compare it with the literature in \S\ref{app:r_v_eval}, describe our results for the extended DESILS-LRG sample in \S\ref{app:extended}, and go over our halo-model calculation for $P_{ge}$ in detail in \S\ref{app:halo_model}.

\section{Evaluating $r_v$.}\label{app:r_v_eval}

In our work we used the surrogate field framework to calculate the correlation efficiency between the true and reconstructed velocity fields $\eta$ which our normalization depends. Here we calculate in a similar manner the correlation coefficient $r_v$ that is commonly used in the literature. The same surrogate construction gives the two variances that go into the calculation of the cross term $\eta_r^{\rm surr}$ of Eq.~(\ref{eq:N_surrogate}), and averaging over the random catalog of one realization, we can calculate: 
\begin{equation}\label{eq:rv_def}
\sigma_{\rm true}^2 \equiv \big\langle (v_r^{\rm true})^2\big\rangle\,,\quad
\sigma_{\rm rec}^2 \equiv \big\langle \hvr^{\,2}\big\rangle\,,\quad
r_v \equiv \frac{\bar\eta_r}{\sigma_{\rm true}\,\sigma_{\rm rec}}\,,
\end{equation}
where $\bar\eta_r=\langle \hvr\,v_r^{\rm true}\rangle$ is the catalog average of the same product that Eq.~(\ref{eq:N_surrogate}) sums object by object.
The surrogate $\hvr$ carries an arbitrary overall scale that cancels in $r_v$, so $\sigma_{\rm rec}$ is fixed only up to that scale, and we quote only $r_v$ here.

We evaluate Eq.~(\ref{eq:rv_def}) on the same surrogate realizations with which we evaluate our normalizations, which carry no kSZ signal.
The true velocity is the linear-theory radial velocity of the same Gaussian field, at the true positions and without the $k_{\rm max}$ cut that the reconstruction applies; and reconstruct both $\hat v_r$ and $v_r^{\rm true}$ in the same way as described in \S\ref{sec:surrogate_normalization}. 
We collect the $r_v$ values we calculate in Table~\ref{tab:rv}; in every case $\sigma_{\rm true}\approx280\,\mathrm{km\,s^{-1}}$ is the linear-theory value on the grid used for the reconstruction.

Simulated samples with a DESILS-LRG mass function and $\sigma_z/(1+z)=0.02$ give $r_v=0.29\pm0.019$ for a main-like and $0.29\pm0.020$ for an extended-like selection~\cite{Guachalla:2023lbx,Hadzhiyska:2023nig}, rising to $0.30\pm0.018$ and $0.31\pm0.019$ once redshift-space distortions are included (which is about half of what the same simulations give for spectroscopic redshifts $r_v\sim0.7$).
Our values sit in the same band.

The two calculations are not the same measurement.
Their true velocity is an $N$-body simulation of host-halo velocities, whose small-scale variance is absent from the linear $\sigma_{\rm true}$ we calculate, and our photometric scatter is $0.024$ and $0.027$ rather than their single $0.02$.
We cut the reconstruction at $k_{\rm max}=0.05\,\mathrm{Mpc}^{-1}$ and apply the same filter $U(k)$ that we define in Eq.~\eqref{eq:vhat_flat_pge} on larger scales, where they smooth with a $12.5\,h^{-1}$Mpc Gaussian; thus, the two are close in scale and different in shape.
We leave a fuller treatment of $\eta$ and $r_v$, including their dependence on the filter, the scale cut, and on the assumed cosmology, to separate work.

\begin{table}[t]
\caption{Velocity--reconstruction correlation coefficient $r_v$ of Eq.~(\ref{eq:rv_def}), on the night coadds, with $\sigma_z/(1+z)=0.024$ (main) and $0.027$ (extended).
The $\pm$ is the scatter among the $300$ (NGC), $200$ (SGC) and $500$ (SPT-3G) realizations, not the uncertainty on the mean, which is smaller by $\sqrt{n}$.}
\label{tab:rv}
\begin{tabular}{@{}l @{\hspace{6mm}} c @{\hspace{6mm}} c@{}}
\toprule
Footprint & main & extended \\
\midrule
ACT DR6, NGC & $0.324\pm0.016$ & $0.298\pm0.015$ \\
ACT DR6, SGC & $0.330\pm0.012$ & $0.305\pm0.012$ \\
SPT-3G, SGC  & $0.310\pm0.018$ & $0.286\pm0.020$ \\
\bottomrule
\end{tabular}
\end{table}

\section{The extended selection}\label{app:extended}

\begin{figure*}[t!]
\centering
\includegraphics[width=\textwidth]{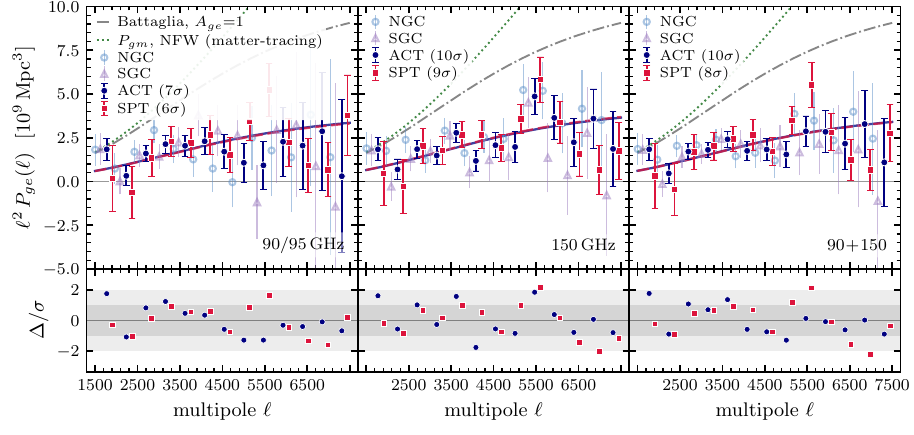}
\vspace*{-27pt}
\caption{Measured $P_{ge}$ bandpowers for the DESILS-LRG extended sample on the night maps, in the conventions of Figure~\ref{fig:bandpowers}: $\sim$\,$90/95$\,GHz (left), $150$\,GHz (centre), and the $90+150$\,GHz combination (right); amplitudes in Table~\ref{tab:detections}.}
\label{fig:bandpowers_ext}
\end{figure*}

\begin{figure*}[t!]
\centering
\includegraphics[width=\textwidth]{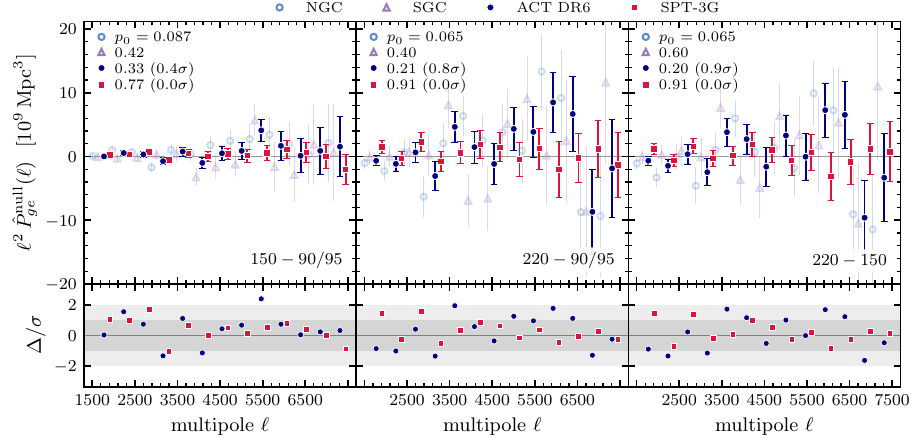}
\vspace*{-21pt}
\caption{Frequency-difference null bandpowers for the DESILS-LRG extended sample on the night maps, at the adopted per-footprint selection, in the conventions of Figure~\ref{fig:freqnull}: the $150-90/95$, $220-90/95$ and $220-150$ differences (Table~\ref{tab:nulls}).}
\label{fig:freqnull_ext}
\end{figure*}

In this section, we briefly describe our results for the extended selection of DESILS-LRG, which we largely omitted in the main part of this paper. 
The ``extended'' DESILS-LRG selection of~Ref.~\cite{Zhou:2023gji} adds fainter galaxies to the main selection, which it contains, at $2.3$ times the comoving number density, $\bar n_g(z_*)=3.46\times10^{-4}$ against $1.50\times10^{-4}\,\mathrm{Mpc}^{-3}$, and a photometric scatter $\sigma_z=0.027$ against $0.024$.
Its hosts are less massive: in our analysis, the occupation matched to its number density gives a galaxy bias $b_g=1.74$ and a satellite fraction $0.094$, compared to $2.02$ and $0.073$ for the main selection (Appendix~\ref{app:halo_model}).
We measure it in the same way throughout, at the same sky selections and binning, in the extended-sample rows of Tables~\ref{tab:detections}, \ref{tab:detections_rvir} and \ref{tab:nulls} and show these results in Figures~\ref{fig:bandpowers_ext} and \ref{fig:freqnull_ext}.

The two surveys agree on the extended selection as they do on the main selection, with \Npairsw{} ACT--SPT-3G channel pairs running to $1.3\sigma$ against $1.9\sigma$ for the main selection, and SPT-3G reaching $8.2\sigma$ on its $95+150$ combination against $8.4\sigma$ for the main selection.
On the $90+150$ GHz combination of both ACT caps, the extended selection returns $10.2\sigma$ against $11.7\sigma$, although it carries $2.6$ times the galaxies; combining SPT-3G with the northern cap gives $\bksz=0.384\pm0.034$ and $11.3\sigma$. 

We associate the lower detection significance with the amplitudes that we measure to our treatment of this selection being suboptimal compared to the main sample, both due to the presence of lower-mass halos than the main sample and larger photo-$z$ errors. We apply no cut on photometric-redshift quality, so we keep the objects with the largest redshift errors (Section~\ref{sec:data_gal}); and we do not weigh the halos by their mass, so we optimize neither in this work.

Our $P_{ge}$ templates account for the lower halo masses of the extended sample, albeit only through the number density matched to that of the extended sample. Since both main and extended samples are fitted to their own template, matched to their own number density, we expect the two amplitudes to agree if the template is equally accurate for both.
At NILC on the night maps, the combined-cap amplitudes are $\bksz=0.351\pm0.035$ for the extended selection and $0.384\pm0.032$ for the main selection; therefore, the extended measurement sits somewhat further below its template than the main measurement does.
Stronger feedback in the lower-mass hosts of the extended selection could produce that difference, and so could an error in our modeling of $P_{ge}$. Here, we refrain from exploring the difference between our fits to the main and extended samples, which we leave to future work. 

All $\NnullExtNight$ null tests pass on the extended catalog for ACT night coadds, with the smallest $p_0$ value at $0.06$.
On the day$+$night coadds \NpassExtDnw{} of the \NnullExtDnw{} extended tests pass; the exception, and the only probability below $0.05$ anywhere in the $\NnullAll$ tests of Table~\ref{tab:nulls}, is the extended northern $220-150$\,GHz difference at $p_0=0.03$, which gives $0.06$ on the night maps.

Evaluated with the mass function of our halo model, the halo occupation Hadzhiyska, Ferraro \emph{et al.} fit to the ACT~DR6 CMB lensing of the extended sample gives a mean host mass of $13.07$~\cite{Hadzhiyska:2025mvt}.
Adopting that occupation as published, the two ACT caps combined place $24.8\pm2.4\%$ of the cosmic baryon budget of the halo within the virial radius, SPT-3G places $24.6\pm3.0\%$, and SPT-3G with the northern cap $25.3\pm2.2\%$.
Matching the occupation to the measured number density instead raises the mean host mass to $13.12$ and lowers the gas fraction measurements to $22.7\pm2.2\%$, $22.6\pm2.8\%$, and $23.2\pm2.0\%$.
The two surveys agree to $0.05\sigma$ under either occupation, and the occupation moves the extended gas fraction by two percentage points against six for the main selection, because the measured number density of the extended selection is closer to what the published occupation predicts.
Hadzhiyska, Ferraro \emph{et al.} quote no gas fraction for the extended sample, reporting instead a modest shortfall in the recovered baryon fraction at large radii, which they attribute to statistical fluctuations, to the larger photometric noise of that selection, or to gas no longer bound to the halo~\cite{Hadzhiyska:2025mvt}.

\section{The halo-model template}\label{app:halo_model}

For the halo model occupation, we consider the 5-parameter HOD model used widely for the DESI analysis~\citep{Zheng:2004id,Yuan:2023ezi}, and we use the DESILS-LRG posterior of Hadzhiyska, Ferraro \emph{et al.}~\cite{Hadzhiyska:2025mvt} to set its parameters. In our model, the electrons follow the Battaglia AGN-feedback fit~\cite{Battaglia:2016xbi}, which we truncate where the enclosed gas reaches the cosmic baryon budget of the halo, a scheme that Qu \emph{et al.}~\cite{Qu:2026zyh} also use (although note that we use $M_{\rm vir}$ in place of $m_{200c}$ for our analysis). 

The $P_{ge}$ template is the sum of one- and two-halo terms:
\begin{equation}\label{eq:Pge_template}
P_{ge}(k,z_*) = P_{ge}^{\mathrm{1h}}(k,z_*) + P_{ge}^{\mathrm{2h}}(k,z_*)\,,
\end{equation}
where the first carries the gas profile of the host halos, which is the small-scale signal we measure. In our analysis, $\bksz$ multiplies both terms. We measure $P_{ge}$ over $\ell\in[1500,7500]$ (or approximately within $k\in[0.56,2.80]{\rm Mpc}^{-1}$) where the one-halo term dominates. We found that our results are insensitive to $\bksz$ multiplied by either both or only the one-halo term. 

The template is computed at $z_*=0.734$ on a grid of $80$ halo masses spanning $2\times10^{10}$--$10^{17}\,M_\odot$, with a Sheth--Tormen mass function and bias~\cite{Sheth:1999su} built on the linear power spectrum of a Planck 2018$+$BAO cosmology~\cite{Planck:2018vyg}.
Grid masses are virial masses, defined by the virial overdensity $\Delta_{\rm vir}(z)$ with respect to the critical density; each is converted to $M_{200c}$ and $R_{200c}$ by rescaling an NFW profile~\cite{Navarro:1995iw} with a Duffy virial concentration~\cite{Duffy:2008pz}.

The gas density follows the Battaglia AGN-feedback fit~\cite{Battaglia:2016xbi}:
\begin{equation}\label{eq:battaglia}
\rho_{\rm gas}(x) = \rho_0\,\Big(\frac{x}{x_c}\Big)^{\gamma}
   \left[1+\Big(\frac{x}{x_c}\Big)^{\alpha}\right]^{-(\beta+\gamma)/\alpha} f_b\,\rho_{\rm cr}(z)\,,
\end{equation}
with $x=r/R_{200c}$, $f_b=\Omega_b/\Omega_m$, and $x_c=0.5$, $\gamma=-0.2$ held fixed. The outer exponent is printed differently in the published and arXiv versions of~\cite{Battaglia:2016xbi}; the form above, in which $\beta$ is the asymptotic slope as that paper states, is the one that reproduces its fitted profiles. The parameter $\rho_0$ and the two exponents $\alpha$ and $\beta$ are power laws in $M_{200c}$ and $(1+z)$ with the AGN-feedback coefficients of that reference.
For each halo, the enclosed gas mass is tabulated out to $100\,$Mpc, and the truncation radius $r_t$ is solved from $M_{\rm gas}(<r_t)=f_b\,M_{\rm vir}$.
The sharp cut is then replaced by the window
\begin{equation}\label{eq:taper}
W(r) = \tfrac{1}{2}\left\{1-\tanh\!\left[\frac{r-r_t'}{\Delta r}\right]\right\}\,,
\qquad \Delta r = 0.25\,R_{200c}\,,
\end{equation}
whose centre $r_t'$ is solved again for each halo so that the tapered profile again encloses $f_b\,M_{\rm vir}$.
The electron form factor $u_e(k)$ is the spherical Fourier transform of the tapered profile, divided by the gas mass that the profile encloses, so $u_e\to1$ as $k\to0$.

The cumulative gas fraction of the template is: 
\begin{equation}
    g(R)=M_{\rm gas}(<R)/(f_b M_{\rm vir})\,,
\end{equation} 
equal to unity at $r_t$ by the construction above, so a fitted amplitude reads as: 
\begin{equation}
    f_{\rm gas}(<R)=\bksz\,g(R)\,.
\end{equation}
We evaluate $g$ as the ratio of the total enclosed gas to the total budget over the host halos, weighted by $dn/dM\,\langle N_{\rm gal}\rangle(M)$ at $z_*$. For the main selection, we get $r_t=1.8\,r_{\rm vir}$ and $g(r_{\rm vir})=0.636$.

For our results in Section~\ref{ssec:signal}, we also replace the fixed profile above with a four-parameter generalized NFW normalized by the same truncation, fitted to the $90/95+150$ bandpowers of each dataset, and calculate from the enclosed gas fractions in a similarly way.

In our HOD model, the centrals and satellites follow:
\begin{align}
N_c(M) &= \tfrac12\,\mathrm{erfc}\!\left[\frac{\log_{10}M_{\rm cut}-\log_{10}M}{\sqrt{2}\,\sigma}\right],
\label{eq:hod_cen}\\[2pt]
N_s(M) &= \begin{cases} N_c(M)\left[\frac{M-\kappa M_{\rm cut}}{M_1}\right]^{\alpha_s}
   & M>\kappa M_{\rm cut},\\
   0 & M\leq\kappa M_{\rm cut}\,,
   \end{cases}
\label{eq:hod_sat}
\end{align}
with satellites tracing the NFW profile of their host.
In Eq.~(\ref{eq:hod_sat}), $\alpha_s$ is the satellite slope, distinct from the gas-profile exponent $\alpha$ of Eq.~(\ref{eq:battaglia}).

As we stated earlier, we take $\sigma$, $\alpha_s$, $\kappa$, and the offset $\log_{10}M_1-\log_{10}M_{\rm cut}$ from the DESILS-LRG posterior of~\cite{Hadzhiyska:2025mvt}, converting its masses from $M_\odot/h$ and solving for $\log_{10}M_{\rm cut}$, shifting $\log_{10}M_1$ with it so that the model reproduces the number densities of Section~\ref{sec:data_gal}:
\begin{equation}\label{eq:hod_params}
\begin{array}{lccccc}
\hline
 & \log_{10}\frac{M_{\rm cut}}{M_\odot} & \log_{10}\frac{M_1}{M_\odot} & \sigma & \alpha_s & \kappa \\
\hline
\text{main}     & 13.038 & 14.418 & 0.133 & 0.848 & 1.245 \\
\text{extended} & 12.729 & 14.434 & 0.108 & 0.642 & 0.941 \\
\hline\\
\end{array}
\end{equation}
The galaxy bias and the satellite fraction follow from this construction. We use $b_g=2.02$ and $f_{\rm sat}=0.073$ for the main sample, and $1.74$ and $0.094$ for the extended sample.

The one-halo term is:
\begin{multline}\label{eq:p1h}
P_{ge}^{\mathrm{1h}}(k) = \int dM\,n(M)\\
  \times\;\frac{N_c+N_s\,u_{\rm NFW}(k|M)}{\bar n_g}\;\frac{M\,u_e(k|M)}{\bar\rho_m}\,,
\end{multline}
with $\bar\rho_m$ the mean matter density and $\bar n_g$ the galaxy number density; the two-halo term multiplies the linear power spectrum by the bias-weighted mass integrals of the same two factors, completed at low mass so that, as $k\to0$, the electrons trace matter and the galaxies carry the bias.

\begin{table*}[t!]
\centering
\caption{Amplitude $\bksz\pm\sigma$ of the measured $\Pge$ relative to the Battaglia-profile prediction, per channel, for the night and the day$+$night coadds.
The $90/95$\,GHz column is $90$\,GHz for ACT and $95$\,GHz for SPT-3G; NILC is night data alone, so its two columns differ only through the analysis selection.
The rows NGC$+$SGC over the two ACT caps, and ACT NGC $+$ SPT-3G, are inverse-variance combinations of the surveys.
}
\vspace{0.8mm}
{\footnotesize
\begin{tabular}{@{}l @{\hspace{2mm}} c@{\hspace{1.2mm}}c @{\hspace{2mm}} c@{\hspace{1.2mm}}c @{\hspace{2mm}} c@{\hspace{1.2mm}}c @{\hspace{2mm}} c@{}}\toprule
 & \multicolumn{2}{c}{$90/95$\,GHz} & \multicolumn{2}{c}{$150$\,GHz} & \multicolumn{2}{c}{NILC} & $90/95+150$ \\
\cmidrule(lr){2-3}\cmidrule(lr){4-5}\cmidrule(lr){6-7}\cmidrule(lr){8-8}
Sample and cap & night & day$+$night & night & day$+$night & night & day$+$night & night \\
\midrule
\multicolumn{8}{@{}l}{\it ACT DR6}\\
main, NGC & $0.406\pm0.068$ & $0.457\pm0.057$ & $0.507\pm0.054$ & $0.512\pm0.049$ & $0.435\pm0.045$ & $0.433\pm0.045$ & $0.442\pm0.049$ \\
main, SGC & $0.432\pm0.067$ & $0.415\pm0.065$ & $0.318\pm0.055$ & $0.291\pm0.061$ & $0.327\pm0.047$ & $0.303\pm0.057$ & $0.373\pm0.049$ \\
main, NGC$+$SGC & $0.419\pm0.048$ & $0.439\pm0.043$ & $0.413\pm0.038$ & $0.426\pm0.038$ & $0.384\pm0.032$ & $0.384\pm0.035$ & $0.408\pm0.035$ \\
\addlinespace[1mm]
extended, NGC & $0.348\pm0.069$ & $0.400\pm0.060$ & $0.479\pm0.055$ & $0.485\pm0.049$ & $0.376\pm0.046$ & $0.383\pm0.046$ & $0.396\pm0.050$ \\
extended, SGC & $0.396\pm0.074$ & $0.409\pm0.077$ & $0.306\pm0.062$ & $0.307\pm0.063$ & $0.318\pm0.053$ & $0.343\pm0.062$ & $0.353\pm0.054$ \\
extended, NGC$+$SGC & $0.370\pm0.050$ & $0.403\pm0.047$ & $0.403\pm0.041$ & $0.417\pm0.039$ & $0.351\pm0.035$ & $0.369\pm0.037$ & $0.376\pm0.037$ \\
\midrule
\multicolumn{8}{@{}l}{\it SPT-3G D1}\\
main, SGC & \multicolumn{2}{c}{$0.439\pm0.059$} & \multicolumn{2}{c}{$0.368\pm0.047$} & \multicolumn{2}{c}{---} & $0.392\pm0.047$ \\
extended, SGC & \multicolumn{2}{c}{$0.366\pm0.058$} & \multicolumn{2}{c}{$0.407\pm0.047$} & \multicolumn{2}{c}{---} & $0.374\pm0.046$ \\
\midrule
\multicolumn{8}{@{}l}{\it ACT NGC $+$ SPT-3G}\\
main & \multicolumn{2}{c}{$0.425\pm0.045$} & \multicolumn{2}{c}{$0.428\pm0.036$} & \multicolumn{2}{c}{---} & $0.416\pm0.034$ \\
extended & \multicolumn{2}{c}{$0.358\pm0.044$} & \multicolumn{2}{c}{$0.438\pm0.036$} & \multicolumn{2}{c}{---} & $0.384\pm0.034$ \\
\bottomrule
\end{tabular}
}
\label{tab:detections}
\end{table*}

\begin{table*}[t!]
\centering
\caption{As Table~\ref{tab:detections}, for the Battaglia template truncated at the virial radius $r_{\rm vir}$ rather than at the gas budget. The template is normalized to the mass they enclose, so they hold the same gas and differ in how it is distributed.}
\vspace{0.8mm}
{\footnotesize
\begin{tabular}{@{}l @{\hspace{2mm}} c@{\hspace{1.2mm}}c @{\hspace{2mm}} c@{\hspace{1.2mm}}c @{\hspace{2mm}} c@{\hspace{1.2mm}}c @{\hspace{2mm}} c@{}}
\toprule
 & \multicolumn{2}{c}{$90/95$\,GHz} & \multicolumn{2}{c}{$150$\,GHz} & \multicolumn{2}{c}{NILC} & $90/95+150$ \\
\cmidrule(lr){2-3}\cmidrule(lr){4-5}\cmidrule(lr){6-7}\cmidrule(lr){8-8}
Sample and cap & night & day$+$night & night & day$+$night & night & day$+$night & night \\
\midrule
\multicolumn{8}{@{}l}{\it ACT DR6}\\
main, NGC & $0.307\pm0.053$ & $0.345\pm0.044$ & $0.375\pm0.040$ & $0.376\pm0.036$ & $0.317\pm0.033$ & $0.318\pm0.033$ & $0.330\pm0.037$ \\
main, SGC & $0.332\pm0.052$ & $0.316\pm0.050$ & $0.230\pm0.041$ & $0.212\pm0.045$ & $0.237\pm0.035$ & $0.220\pm0.043$ & $0.275\pm0.037$ \\
main, NGC$+$SGC & $0.319\pm0.037$ & $0.332\pm0.033$ & $0.304\pm0.029$ & $0.313\pm0.028$ & $0.280\pm0.024$ & $0.281\pm0.026$ & $0.303\pm0.026$ \\
\addlinespace[1mm]
extended, NGC & $0.263\pm0.054$ & $0.308\pm0.047$ & $0.361\pm0.041$ & $0.359\pm0.037$ & $0.282\pm0.035$ & $0.286\pm0.035$ & $0.298\pm0.039$ \\
extended, SGC & $0.313\pm0.060$ & $0.312\pm0.061$ & $0.228\pm0.048$ & $0.229\pm0.048$ & $0.236\pm0.041$ & $0.255\pm0.048$ & $0.269\pm0.042$ \\
extended, NGC$+$SGC & $0.286\pm0.040$ & $0.310\pm0.037$ & $0.304\pm0.031$ & $0.310\pm0.029$ & $0.263\pm0.027$ & $0.275\pm0.028$ & $0.285\pm0.029$ \\
\midrule
\multicolumn{8}{@{}l}{\it SPT-3G D1}\\
main, SGC & \multicolumn{2}{c}{$0.307\pm0.042$} & \multicolumn{2}{c}{$0.251\pm0.033$} & \multicolumn{2}{c}{---} & $0.267\pm0.033$ \\
extended, SGC & \multicolumn{2}{c}{$0.257\pm0.042$} & \multicolumn{2}{c}{$0.284\pm0.034$} & \multicolumn{2}{c}{---} & $0.258\pm0.033$ \\
\bottomrule
\end{tabular}
}
\label{tab:detections_rvir}
\end{table*}

\begin{table*}[t!]
\centering
\caption{Null probabilities $p_0$ for the $\NnullAll$ tests, at the selections and binning of Table~\ref{tab:detections}; the ACT tests are repeated on the night and the day$+$night (``d$+$n'') coadds.
Left: differences of the beam-equalized single-frequency spectra within one cap, in which the blackbody kSZ signal should cancel.  Right: the difference of the two Galactic caps (Section~\ref{ssec:nulls}).}
\vspace{0.8mm}
\begin{tabular}[t]{@{}l @{\hspace{4mm}} c@{\hspace{2.5mm}}c @{\hspace{5mm}} c@{\hspace{2.5mm}}c @{\hspace{5mm}} c@{\hspace{2.5mm}}c@{}}
\multicolumn{7}{@{}l}{\it Frequency difference}\\
\toprule
 & \multicolumn{2}{c}{$150-90/95$} & \multicolumn{2}{c}{$220-90/95$} & \multicolumn{2}{c}{$220-150$}\\
\cmidrule(lr){2-3}\cmidrule(lr){4-5}\cmidrule(lr){6-7}
 & night & d$+$n & night & d$+$n & night & d$+$n \\
\midrule
ACT main, NGC & 0.48 & 0.41 & 0.24 & 0.55 & 0.19 & 0.20 \\
ACT main, SGC & 0.19 & 0.16 & 0.60 & 0.39 & 0.61 & 0.17 \\
ACT extended, NGC & 0.09 & 0.14 & 0.07 & 0.09 & 0.06 & 0.03 \\
ACT extended, SGC & 0.42 & 0.62 & 0.40 & 0.50 & 0.60 & 0.42 \\
\addlinespace[0.7mm]
SPT-3G main, SGC & \multicolumn{2}{c}{0.10} & \multicolumn{2}{c}{0.61} & \multicolumn{2}{c}{0.67} \\
SPT-3G extended, SGC & \multicolumn{2}{c}{0.77} & \multicolumn{2}{c}{0.91} & \multicolumn{2}{c}{0.91} \\
\bottomrule
\end{tabular}%
\hspace{9mm}%
\begin{tabular}[t]{@{}l @{\hspace{4mm}} c@{\hspace{2.5mm}}c @{\hspace{5mm}} c@{\hspace{2.5mm}}c@{}}
\multicolumn{5}{@{}l}{\it Cap difference, ACT (NGC$-$SGC)}\\
\toprule
 & \multicolumn{2}{c}{main} & \multicolumn{2}{c}{extended}\\
\cmidrule(lr){2-3}\cmidrule(lr){4-5}
 & night & d$+$n & night & d$+$n \\
\midrule
$90$\,GHz & 0.83 & 0.61 & 0.68 & 0.61 \\
$150$\,GHz & 0.06 & 0.09 & 0.59 & 0.47 \\
$220$\,GHz & 0.50 & 0.82 & 0.35 & 0.52 \\
NILC & 0.60 & 0.57 & 0.996 & 0.95 \\
\bottomrule
\end{tabular}

\label{tab:nulls}
\end{table*}

\bibliography{refs}

\end{document}